\documentclass[traditabstract]{aa}
\usepackage{epsfig}
\usepackage{natbib}

\def\gtsima{$\; \buildrel > \over \sim \;$}
\def\ltsima{$\; \buildrel < \over \sim \;$}
\def\gtrsim{\lower.5ex\hbox{\gtsima}}
\def\lesssim{\lower.5ex\hbox{\ltsima}}

\begin{document}

\title{The Gaia-ESO Survey: N-body modelling of the Gamma Velorum cluster}

\author{M. Mapelli\inst{1}, A. Vallenari\inst{1},  R. D. Jeffries\inst{2}, E. Gavagnin\inst{1,3}, T. Cantat-Gaudin\inst{1}, G. G. Sacco\inst{4}, M. R. Meyer\inst{5}, E. J. Alfaro\inst{6}, M. Costado\inst{6}, F. Damiani\inst{7}, A. Frasca\inst{8}, A. C. Lanzafame\inst{9},  S. Randich\inst{4}, R. Sordo\inst{1}, S. Zaggia\inst{1}, G. Micela\inst{7}, E. Flaccomio\inst{7}, E. Pancino\inst{10,11},  M. Bergemann\inst{12}, A. Hourihane\inst{12}, C. Lardo\inst{13},  L. Magrini\inst{4}, L.~Morbidelli\inst{4}, L. Prisinzano\inst{7}, C.~C. Worley\inst{12}} 
\institute{INAF-Osservatorio Astronomico di Padova, Vicolo dell'Osservatorio 5, I--35122, Padova, Italy\\ \email{michela.mapelli@oapd.inaf.it}
\and
Astrophysics Group, Keele University, Keele, Staffordshire ST5 5BG, United Kingdom   
\and
Institute for Computational Science, University of Zurich, Winterthurerstrasse 190, CH-8057 Zurich, Switzerland
\and
INAF - Osservatorio Astrofisico di Arcetri, Largo E. Fermi 5, I-50125, Florence, Italy
\and
Institute for Astronomy, ETH Zurich, Wolfgang-Pauli-Strasse 27, CH-8093, Zurich, Switzerland
\and
Instituto de Astrof\'{i}sica de Andaluc\'{i}a-CSIC, Apdo. 3004, 18080, Granada, Spain
\and
INAF - Osservatorio Astronomico di Palermo, Piazza del Parlamento 1, 90134, Palermo, Italy
\and
INAF - Osservatorio Astrofisico di Catania, via S. Sofia 78, 95123, Catania, Italy
\and
Dipartimento di Fisica e Astronomia, Sezione Astrofisica, Universit\`{a} di Catania, via S. Sofia 78, 95123, Catania, Italy
\and
INAF - Osservatorio Astronomico di Bologna, via Ranzani 1, 40127, Bologna, Italy
\and                          
ASI Science Data Center, Via del Politecnico SNC, 00133 Roma, Italy
\and
Institute of Astronomy, University of Cambridge, Madingley Road, Cambridge CB3 0HA, United Kingdom
\and
Astrophysics Research Institute, Liverpool John Moores University, 146 Brownlow Hill, Liverpool L3 5RF, United Kingdom
}
\titlerunning{N-body modelling of the Gamma Velorum cluster}
 
\authorrunning{Mapelli et al.}
 
\abstract{
The Gaia-ESO Survey has recently unveiled the complex kinematic signature of the Gamma Velorum cluster: this cluster is composed of two kinematically distinct populations (hereafter, population A and B), showing two different velocity dispersions and  a relative $\sim{}2$ km s$^{-1}$ radial velocity (RV) shift. In this paper, we propose that the two populations of the Gamma Velorum cluster originate from two different sub-clusters, born from the same parent molecular cloud. We investigate this possibility by means of direct-summation N-body simulations. Our scenario is able to reproduce not only the RV shift and the different velocity dispersions, but also the different centroid ($\sim{}0.5$ pc), the different spatial concentration and the different line-of-sight distance ($\sim{}5$ pc) of the two populations. The observed $1-2$ Myr age difference between the two populations is also naturally explained by our scenario, in which the two sub-clusters formed in two slightly different star formation episodes. Our simulations suggest that population~B is strongly supervirial, while population~A is close to virial equilibrium. We discuss the implications of our models for the formation of young star clusters and OB associations in the Milky Way.
}

\maketitle

%

\section{Introduction}~\label{sec:intro}
The ongoing Gaia-ESO Survey (GES) at the VLT is a gold-mine of knowledge concerning the formation and evolution of young star clusters (YSCs) and OB associations. The aim of the GES, which began on December 31 2011 and will be completed in $\sim{}5$ years, is to obtain high quality, uniformly calibrated spectra of $>10^5$ stars in the Milky Way (\citealt{Gilmore2012}; \citealt{RandichGilmore2013}). The GES targets include a significant number of YSCs. 
The radial velocity and the chemical composition is being derived for a large number of YSC members with high accuracy. This information, when combined with data from the Gaia mission (\citealt{Perryman2001}), will enable reconstruction of the full three-dimensional spatial distribution and kinematics of several YSCs.

The first GES YSC target was the Gamma Velorum cluster, a marginally bound cluster of low-mass pre-main sequence (PMS) stars, surrounding the massive Wolf-Rayet (WR) - O binary system named $\gamma{}^2$ Velorum (HD~68273, WR11; \citealt{Smith1968,Schaerer1997}).  The Wolf-Rayet and O star components of  $\gamma{}^2$ Velorum have current masses $\sim{}9\pm{}2$ and $30\pm{}2$ M$_\odot{}$, respectively \citep{Demarco1999}, which implies that the initial masses were  $\sim{}35$ and 31 M$_\odot{}$, respectively \citep{Eldridge2009}. $\gamma{}^2$ Velorum is the most massive member of the common proper motion Vela OB2 association, composed of 93 early-type stars, spread over 100 square degrees (\citealt{deZeeuw1999}). The Gamma Velorum cluster was firstly identified by \cite{Pozzo2000}, on the basis of the strong X-ray emission of the PMS stars, and then further investigated by  \cite{Jeffries2009}.  The distance of the Gamma Velorum cluster is about 350~pc (\citealt{Pozzo2000,Vanleeuwven2007,Millour2007,North2007}). The PMS stars in the Gamma Velorum cluster have an estimated age $\sim{}5-10$ Myr (or even more), which is only marginally consistent with the age of the $\gamma{}^2$ Velorum binary ($\sim{}5.5$ Myr, \citealt{Eldridge2009}). \cite{Jeffries2014} suggest, on the basis of the Lithium (Li) depletion observed among its M-dwarf members, that the Gamma Velorum cluster is older than 10 Myr. 

The GES data (\citealt{Jeffries2014}) show that the Gamma Velorum cluster has a complex kinematic structure: it is composed of two kinematically distinct populations (named population A and B by \citealt{Jeffries2014}). Population~A (hereafter pop.~A) has a $2.15\pm{}0.48$ km s$^{-1}$ lower radial velocity (RV) than population~B (hereafter pop.~B). The velocity dispersion  of  the two populations is also different: pop.~A (pop.~B) has a best-fitting velocity dispersion $\sigma{}_{\rm A}=0.34\pm{}0.16$ km s$^{-1}$ ($\sigma{}_{\rm B}=1.60\pm{}0.37$ km s$^{-1}$).

 The colour-magnitude diagram suggests that either pop.~A is younger than pop.~B by $0.4\pm{}0.6$ Myr  (if the two populations are at the same distance), or pop.~A and pop.~B are coeval but pop.~A is closer to us by $4\pm{}5$ pc. The spectral analysis indicates a possible age difference, but in the opposite sense: pop.~A appears to be marginally more Li depleted than pop.~B, which implies that pop.~A might be $1-2$ Myr older.  The age difference derived from Li depletion, combined with the analysis of the colour-magnitude diagram, hints that pop.~A might be closer to us than pop.~B by a few parsecs.

Finally, the spatial distribution also suggests differences between  pop.~A and B: the former is slightly more concentrated than the latter (see Figure 11 of \citealt{Jeffries2014}), and the centroids of the two populations are offset by $4.4\pm{}3.0$ arcmin ($0.4\pm{}0.3$ pc for an assumed distance of $350$ pc). The centroid of pop.~A is almost coincident with the position of  $\gamma{}^2$ Velorum, indicating a connection between pop.~A and the WR-O binary system.

\cite{Jeffries2014} propose three possible scenarios for the formation of the Gamma Velorum cluster. In the first scenario, pop.~A and pop.~B are the remnants of a larger cluster. After the evaporation of the parent molecular gas, only pop.~A remained bound and in virial equilibrium, while pop.~B became unbound. This scenario fails to explain the difference in RV between the two populations and the different Li depletion. In the second scenario, $\gamma{}^2$ Velorum formed in a supervirial OB association. Pop.~A represents the component of the initial OB association which was gravitationally captured by  $\gamma{}^2$ Velorum (e.g. \citealt{Parker2014}), while pop.~B remained unbound. This second scenario explains the velocity differences, but cannot easily explain the age differences between the two populations. In the third scenario, favoured by \cite{Jeffries2014}, pop.~A formed in a denser region, while pop.~B formed in a less dense region: pop.~A is still marginally bound, while pop.~B expands in the Vela OB2 association. This scenario could explain an age difference between the two populations, as well as a different spatial position and a different RV, because pop.~A and pop.~B formed in two slightly different star formation episodes.

In this paper, we focus on the third aforementioned scenario, investigating its implications in detail, by means of an N-body model of the Gamma Velorum cluster. 
The main idea is the following: simulations of turbulent molecular clouds suggest that star clusters form from the merger of sub-clusters (\citealt{Bonnell2003}; \citealt{Bate2009}; \citealt{Bonnell2011}; \citealt{Girichidis2011}; \citealt{Kruijssen2012}). Furthermore, observations of star clusters embedded in molecular clouds indicate that the process of star formation is highly sub-structured (e.g. \citealt{Cartwright2004,Gutermuth2009,Sanchez2009,Andre2010, Demarchi2013,Gouliermis2014,Kuhn2014,Kirk2014}). The merger of sub-clusters can also explain some intriguing dynamical properties of YSCs (e.g. a fast mass segregation, \citealt{Parker2011, Fujii2012,Parker2014}). Thus, we propose that the two populations of the Gamma Velorum cluster originate from two different sub-clusters, born in the same parent molecular cloud. The velocity shift between two sub-clusters is expected to be of the same order of magnitude as the turbulent velocity in the parent molecular cloud.  The observed turbulent motions of Galactic molecular clouds are of the order of $\sim{}1-2$ km s$^{-1}$ (e.g. \citealt{Mckee2007} and references therein), significantly larger than the sound speed of cold gas ($\sim{}0.2$ km s$^{-1}$). Furthermore, stars in the Orion Nebula Cluster (\citealt{Tobin2009}) and molecular gas in star-forming regions such as Perseus and Ophiuchus (\citealt{Andre2007,Kirk2007,Rosolowsky2008}) exhibit large-scale linear velocity gradients ($\lesssim{}1$  km s$^{-1}$ pc$^{-1}$). Velocity gradients ($\sim{}0.2-2$ km s$^{-1}$ pc$^{-1}$) have also been found in simulations of molecular clouds and YSC formation (e.g. \citealt{Offner2009}). Such velocity offsets would naturally explain the observed RV difference of 2 km s$^{-1}$ between pop.~A and pop.~B.  We investigate this possibility by means of direct-summation N-body simulations.

This paper is organized as follows. In Section~2, we discuss the simulation method. In Section~3, we present our results. In Section~4, we discuss the main advantages and drawbacks of our simulation method, and we propose future simulations and observational tests. 





\begin{table*}
\begin{center}
\caption{Initial conditions of the $N-$body simulations.}
 \leavevmode
\begin{tabular}{lllllllllllll}
\hline
Run   & $N_{\rm A}$ & $M_{\rm A}$ (M$_\odot$) & $r_{\rm A}$ (pc)  & $Q_{\rm A}$ & $N_{\rm B}$ & $M_{\rm B}$  (M$_\odot$) & $r_{\rm B}$ (pc)  & $Q_{\rm B}$  & $f_{\rm bin}$ & $\Delta{}v$ (km s$^{-1}$) & $D$ (pc)  & $m_1,\,{}m_2$ (M$_\odot$)\\
\hline
Run~1 & 780       & 506                  & 2               & 1.0                & 585       & 271        & 1               & 4.5        & 0.46        & 2.4                      & 5 & $32,\,{}14$ \\ 
Run~2 & 780       & 510                  & 2               & 0.5                & 585       & 271        & 1               & 4.5        & 0.46        & 2.4                      & 5 & $33,\,{}16$ \\ 
Run~3 & 780       & 514                  & 2               & 2.0                & 585       & 271        & 1               & 4.5        & 0.46        & 2.4                      & 5 & $39,\,{}6$ \\ 
Run~4  & 780       & 513                  & 2               & 3.0                & 585       & 271        & 1               & 4.5        & 0.46        & 2.4              & 5 & $15,\,{}15$ \\ 
Run~5  & 780       & 513                  & 2               & 1.0                & 585       & 276        & 1               & 0.5        & 0.46        & 2.4               & 5 & $32,\,{}14$  \\ 
Run~6  & 780       & 513                  & 2               & 1.0                & 585       & 277        & 1               & 2.0        & 0.46        & 2.4               & 5 & $32,\,{}14$ \\ 
Run~7  & 780       & 513                  & 2               & 1.0                & 585       & 273        & 1               & 12.5        & 0.46        & 2.4               & 5 & $32,\,{}14$  \\ 
Run~8  & 780       & 506                  & 2               & 1.0                 & 585       & 271        & 1               & 4.5        & 0.46        & 1.5              & 5 & $32,\,{}14$  \\ 
Run~9  & 780       & 506                  & 2               & 1.0                 & 585       & 271        & 1               & 4.5        & 0.46        & 2.0              & 5  & $32,\,{}14$ \\ 
Run~10 & 780       & 506                  & 2               & 1.0                 & 585       & 271        & 1               & 4.5        & 0.46        & 3.0              & 5  & $32,\,{}14$ \\ 
Run~11 & 780       & 506                  & 2               & 1.0                 & 585       & 271        & 1               & 4.5        & 0.46        & 3.5              & 5  & $32,\,{}14$ \\ 
Run~12 & 156       & 99                   & 2               & 1.0                 & 117       & 56         & 1               & 4.5        & 0.46        & 2.4              & 5  & $5,\,{}4$  \\ 
Run~13 & 156       & 97                   & 2               & 4.0                 & 117       & 48         & 1               & 8.0          & 0.46        & 2.2              & 5  & $3,\,{}3$    \\ 
Run~14 & 780       & 506                  & 2               & 1.0                & 585       & 271         & 1               & 4.5        & 0.46        & 2.4              & 10  & $32,\,{}14$ \\ 
Run~15 & 780       & 512                  & 2               & 1.0                & 585       & 295         & 1               & 4.5        & 0.0         & 2.4              & 5 & $--$ \\ 
Run~16 & 800       & 512                  & 2               & 1.0                & 560       & 271         & 1               & 4.5        & 0.75          & 2.4              & 5 & $57,\,{}32$ \\
Run~17 & 780       & 491                  & 1               & 1.0                & 585       & 271         & 1               & 4.5        & 0.46        & 2.4              & 5 & $8,\,{}1$ \\ 
Run~18 & 780       & 506                  & 2               & 1.0                & 585       & 272         & 2               & 4.5        & 0.46        & 2.4              & 5 &  $32,\,{}14$\\ 
Run~19 & 1040      & 677                  & 2               & 1.0                 & 585       & 271        & 1              & 4.5        & 0.46        & 2.4              & 5  & $56,\,{}35$\\
Run~20 & 780       & 506                  & 2               & 1.0                &  780       & 359        & 1              & 4.5        & 0.46        & 2.4              & 5 &  $32,\,{}14$\\
Run~21 & 780       & 506                  & 2               & 1.0                & 585       & 271        & 1               & 4.5        & 0.46        & -2.4                      & 5 & $32,\,{}14$ \\ 
\noalign{\vspace{0.1cm}}
\hline
\end{tabular}
\begin{flushleft}
\footnotesize{Run (column 1): identifying name of the run; $N_{\rm A}$ (column 2):  number of particles in the first cluster (identified with pop.~A); $M_{\rm A}$ (column 3): initial mass of the first cluster; $r_{\rm A}$ (column 4):  scale radius of the first cluster; $Q_{\rm A}$ (column 5): virial ratio of pop.~A; $N_{\rm B}$ (column 6):  number of particles in the second cluster (identified with pop.~B); $M_{\rm B}$ (column 7): initial mass of the second cluster; $r_{\rm B}$ (column 8):  scale radius of the second cluster; $Q_{\rm B}$ (column 9): virial ratio of pop.~B; $\Delta{}v$  (column 10): initial velocity shift between the centres of mass of the two clusters;  $D$  (column 11): initial distance between the centres of mass of the two clusters; $m_1,\,{}m_2$  (column 12): initial mass of the members of the most massive binary system in each run. The 21 runs listed in this Table are a selected sample among all the runs we performed. They represent the runs  that best match the observations (runs~1, 2, 13 and 21), and/or that are important to understand the influence of various parameters on the results.\label{tab:tab1}}
\end{flushleft}
\end{center}
\end{table*}
\section{Methods and simulations}~\label{sec:nbody}
In this paper, we simulate the collision between two different sub-clusters, by means of direct-summation N-body simulations. The simulations were performed with the {\sc starlab} public software environment \citep{Portegies2001}, in the version modified by Mapelli et al. (2013; see also \citealt{MapelliBressan2013}). {\sc starlab} integrates the dynamical evolution of a star cluster, by means of a predictor-corrector Hermite scheme (implemented in the {\sc kira} routine), and follows the stellar and binary evolution of the star cluster members (implemented in {\sc seba}).

Initially, the two sub-clusters are described as two Plummer spheres (\citealt{Plummer1911}). The single stars and the primary members of binaries were randomly generated following a Kroupa initial mass function (IMF, \citealt{Kroupa2001}) with minimum and maximum mass 0.1 and 150 M$_\odot$, respectively. The secondary members of the binaries were randomly generated from a uniform distribution between 0.1 M$_\odot$ and $m_1$ (where  $m_1$ is the mass of the primary member). 
We enable stellar and binary evolution at solar metallicity (consistent with \citealt{Spina2014}), as described in \cite{MapelliBressan2013}. 
In the following, we call pop.~A and pop.~B the population of the first and the second simulated YSCs, respectively.

\begin{figure}
\center{{
\epsfig{figure=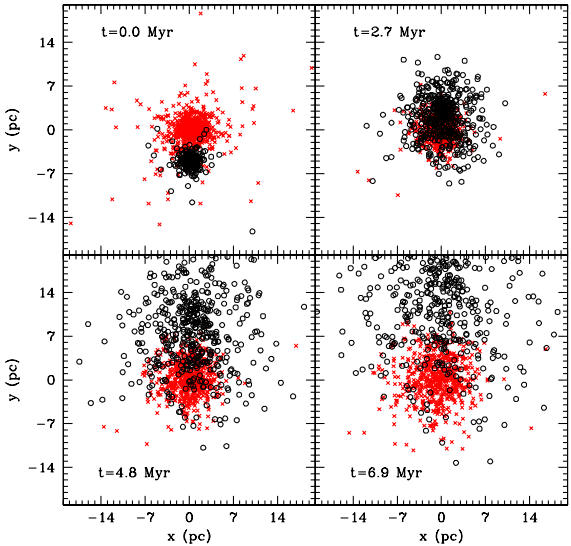,width=8cm} 
}}
\caption{\label{fig:fig1}
Position of the simulated stars in the $xy$ plane at $t=0,\,{}2.7,\,{}4.8$ and 6.9 Myr in run~1. Red crosses: cluster 1 (corresponding to population~A), black open circles: cluster 2 (corresponding to population~B). The centre of the frame is the centre of mass of cluster 1 (pop.~A).
}
\end{figure}
We ran several simulations with different initial conditions. The most relevant runs and the corresponding parameters are summarized in Table~\ref{tab:tab1}. The most important initial parameters appear to be (i) the number of particles in each star cluster ($N_{\rm A}$ and $N_{\rm B}$, respectively) and the corresponding initial masses ($M_{\rm A}$ and $M_{\rm B}$, respectively); (ii) the Plummer scale radius of each star cluster ($r_{\rm A}$ and $r_{\rm B}$, respectively); (iii) the relative velocity ($\Delta{}v$) and  the initial distance  ($D$) between the centres-of-mass of the two clusters; (iv) the binary fraction $f_{\rm bin}$; (v) the deviation of the star cluster from the virial equilibrium. The last condition is particularly important, because several observed YSCs show indication of non-virial velocities (e.g. \citealt{Cottaar2012}). 
 To quantify the deviation from virial equilibrium, we define the  $Q$ parameter as $Q\equiv{}K/|W|$ (where $K$ and $W$ are the kinetic and the potential energy of the system, respectively). For a system in virial equilibrium $Q=0.5$. Systems with $Q>0.5$ (i.e. supervirial systems) might be the outcome of the evaporation of the parent gas. In fact, the gas component is not included in our models, but simulating a stellar population with initial velocities above the virial value mimics the fact that the potential well changed very fast in the recent past, i.e. that the parent gas evaporated rapidly.

 We test two extreme values for the initial mass: a total mass (pop.~A+pop.~B) of $\sim{}150$ M$_\odot{}$ (runs~12 and 13) and of $\sim{}800$ M$_\odot{}$ (the other runs). The mass of the clusters simulated in runs~12 and 13 is similar to the current total mass of the Gamma Velorum cluster (where for current total mass we mean the mass within the 0.9 deg$^2$ area surveyed by GES), while a total mass of $\sim{}800$ M$_\odot{}$ is likely closer to the initial mass of the system. The last column of Table~\ref{tab:tab1} shows the most massive primordial binary system in each run. A binary as massive as $\gamma{}^2$ Velorum is likely to form only in the most massive systems amongst those simulated here ($\sim{}800$ M$_\odot{}$). The large scatter in the values of $M_{\rm A}$ and $M_{\rm B}$  listed in Table~\ref{tab:tab1} (given a fixed value of $N_{\rm A}$ and $N_{\rm B}$, respectively) is due to stochastic fluctuations, because we randomly generate the masses of star particles according to a Kroupa IMF.  

In run~21 the two sub-clusters do not collide. They are initially located at $D=5$ pc (Table~\ref{tab:tab1}), and their separation increases (negative $\Delta{}v$). This run has been performed to check whether the properties of the Gamma Velorum cluster are consistent with a mere superposition between two sub-clusters (without physical interaction).

\section{Results}~\label{sec:results}
\begin{figure*}
\center{{
\epsfig{figure=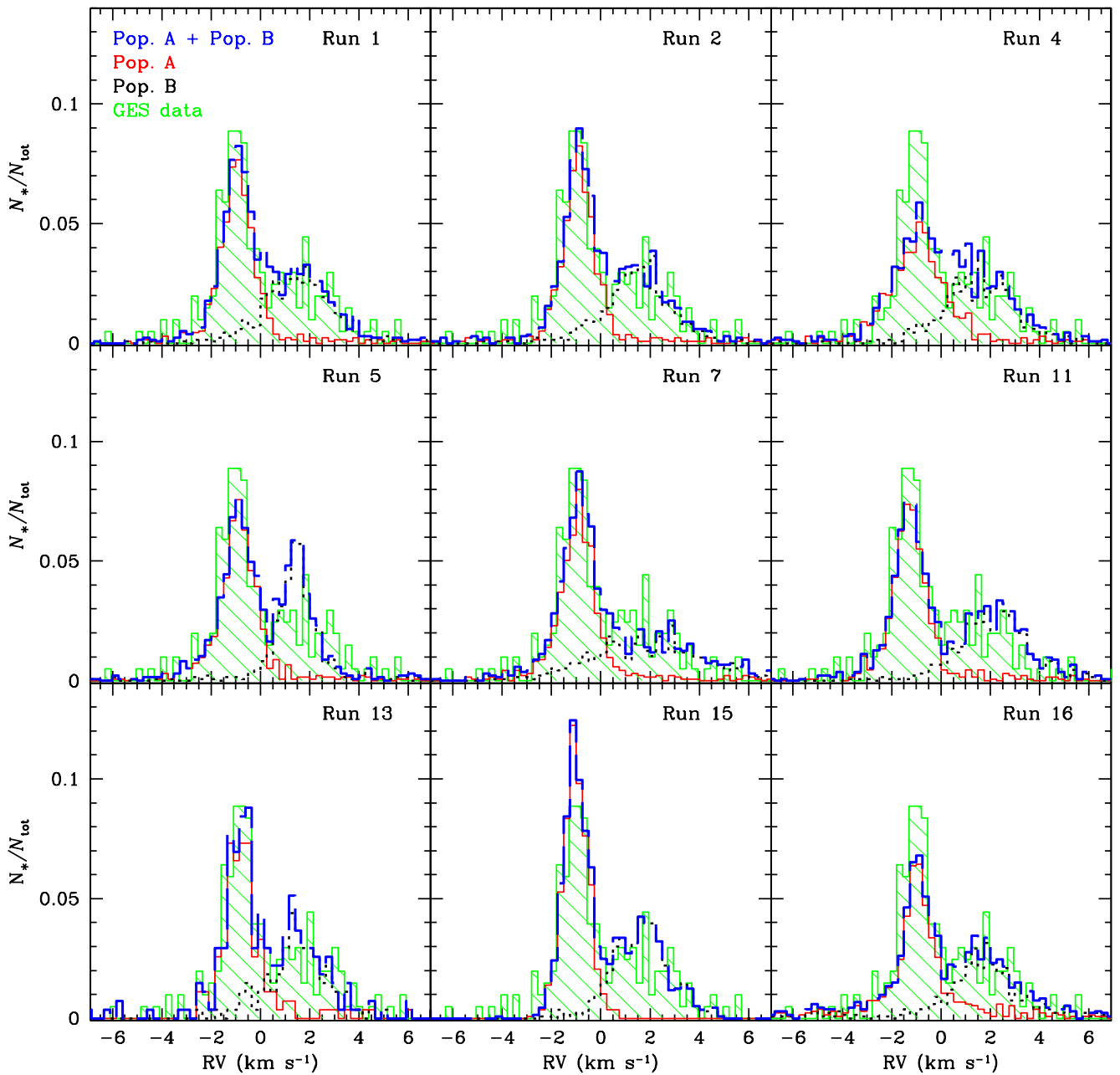,width=18cm} 
}}
\caption{\label{fig:fig2}
From top to bottom and from left to right: distribution of the radial velocities (RVs) of stars in run~1, 2, 4, 5, 7, 11, 13, 15 and 16 at $t=4.8$ Myr (see Table~\ref{tab:tab1}). The simulated RVs include randomly generated observational uncertainties. Solid red line: pop.~A (cluster 1); dotted black line: pop.~B (cluster 2); dashed blue line: sum of the two components. We define RV as the velocity along the $y$ axis (i.e. we assume that the line-of-sight coincides with the direction that maximizes the velocity shift between the two clusters). Green shaded histogram: observed RV distribution (the same as in Figure~6 of \citealt{Jeffries2014}).
}
\end{figure*}
We simulate the collision between two different YSCs, during and after the first approach. Figure~\ref{fig:fig1} shows the time evolution of run~1 in the $xy$ plane. 
In all runs, the initial velocity shift is entirely along the  $y$ axis. In run~1, the two clusters are initially at a distance of 5 pc along the $y$ axis (which is less than the typical radius of a giant molecular cloud in the Milky Way, e.g. \citealt{Dame2001}). They collide at time $t\sim{}2.5$ Myr. The YSCs simulated in the other runs have similar evolution, but the collision time is different, since it depends on the relative initial distance, velocity and masses.

 Detailed information about all runs (including those shown in Figures~\ref{fig:fig1} and ~\ref{fig:fig2}) can be found in Table~\ref{tab:tab2}. In calculating the RVs in Table~\ref{tab:tab2} (as well as in Figure~\ref{fig:fig2}), we have assumed that the line-of-sight is along the $y$ axis. Thus, the RV is the velocity along the $y$ axis, i.e. the axis along which we generated the velocity shift $\Delta{}v$ between the centre-of-mass of the two clusters\footnote{Without this requirement, the probability of observing the two populations nearly superimposed would be much smaller, even if this is would be compensated by a less stringent requirement on the line-of-sight direction. See Section~\ref{sec:LOS} for details.}. Therefore, the plane of the sky is the $xz$ plane of the simulations.

 In Table~\ref{tab:tab2}, we consider as reference times $t\sim{}2.5,\,{}5$ and $7$ Myr since the beginning of the simulations. We stress that we select these three reference snapshots, since the radial distribution of the simulated stars, and especially the shift along the line-of-sight between pop.~A and pop.~B, are not consistent with the observations if $t<<2.5$ Myr or $t>>7.5$ Myr (see Section~\ref{sec:shift}). On the other hand, the RV distribution does not change significantly at different times (see Table~\ref{tab:tab2}), provided that  $t\ge{}2.5$ Myr (i.e. provided that the two sub-clusters already collided). Thus, the main constraints on the time elapsed since the collision come from the radial distribution, while the RV distribution is less affected. Finally, we recall that the time indicated in Table~\ref{tab:tab2} does not necessarily correspond to the age of the stars in pop.~A and pop.~B: the tabulated time is the time elapsed since the beginning of the simulations, but the two sub-clusters might have formed before $t=0$. Evaluating the age of the Gamma Velorum cluster, which is rather uncertain (see the Introduction), is beyond the aims of this paper.

\subsection{Radial velocity and proper motions}\label{sec:RV}
Figure~\ref{fig:fig2} shows the distribution of the RVs for nine selected runs at $t=4.8$ Myr (for other runs, see Table~\ref{tab:tab2}).   The simulated RVs shown in Figure~\ref{fig:fig2} include observational uncertainties. These were randomly drawn from the distribution reported in Figure~2 of \cite{Jeffries2014}. The zero-point of the simulated RVs is the RV velocity of the centre-of-mass of the total system (pop.~A+pop.~B stars).  In Figure~\ref{fig:fig2}, we also show the observed RV distribution (the same as in Figure~6 of \citealt{Jeffries2014}). The observed RV distribution has been shifted to match the simulated ones.


In our simulations, we exactly know which stars belong to the first and to the second cluster, respectively, and we can derive their intrinsic velocity dispersions\footnote{The intrinsic velocity dispersion for each population was derived as the standard deviation from the line-of-sight velocity, i.e. $\tilde{\sigma{}}^2=\sum_i(v_i-\langle{}v\rangle{})^2/(N-1)$ (where $v_i$ are the line-of-sight velocities of each star, $\langle{}v\rangle{}$ is the average line-of-sight velocity and $N$ is the number of particles). To derive $\tilde{\sigma{}}$, we consider only the centre-of-mass motion of binary systems.}. Columns 7 and 8 of Table~\ref{tab:tab2} show the intrinsic velocity dispersions ($\tilde{\sigma{}}_{\rm A}$ and $\tilde{\sigma{}}_{\rm B}$) of the simulated pop.~A and pop.~B, along the line-of-sight. Column~9 of Table~\ref{tab:tab2} ($\tilde{\Delta{}}$RV) shows the intrinsic velocity shift of the two simulated populations along the line-of-sight, i.e. the difference between the mean velocity of pop.~A and that of pop.~B along the line-of-sight.

On the other hand, we want to fit the total RV distribution of each simulation in the same way as \cite{Jeffries2014}, for a better comparison with the data of the Gamma Velorum cluster. Thus, we apply a fit with two Gaussian components to the total pop.~A+pop.~B distribution, for all runs. The fitting procedure is carried out on binned data, and adopting a procedure equivalent to the one described in section~4 of Jeffries et al. (2014, i.e. statistically correcting for the binary fraction and for the observational uncertainties {\it a posteriori}).  From this fitting procedure, we obtain the velocity dispersions of the two simulated populations along the line-of-sight ($\sigma{}_{\rm A}$ and $\sigma{}_{\rm B}$), the velocity shift of the two simulated populations along the line-of-sight ($\Delta{}$RV), and the fraction of estimated pop.~A members with respect to the total number of simulated stars ($f_{\rm A}$). The results are shown in columns  3, 4, 5 and 6 of Table~\ref{tab:tab2}. 
The intrinsic velocity dispersions generally agree with the ones derived from this fitting procedure, but they might also disagree by a factor of $\sim{}2$ (especially for pop.~B), when the distribution of simulated RVs deviates too much from a Gaussian distribution, and when it is more difficult to disentangle pop.~A from pop.~B.

 We adopt two criteria to check whether a simulation is consistent with the observations of the Gamma Velorum cluster. Namely, we decide that a simulation (at a given time) is not consistent with the observations if (i) at least one of the six reference quantities ($\sigma{}_{\rm A}$,   $\sigma{}_{\rm B}$, $\Delta{}$RV, $f_{\rm A}$, the distance between the centre of mass of pop. A and pop. B along the line-of-sight $\Delta{}y$, and  the distance between the centre of mass of pop. A and  pop. B in the plane of the sky $\Delta{}xz$) differs from the observed one by more than 3 $\sigma$, or (ii)  $P_{\rm KS}<0.05$  (where  $P_{\rm KS}$ is the probability that the chance deviation between the observed RV distribution and the simulated RV distribution is expected to be larger, according to a Kolmogorov-Smirnov test, last column of  Table~\ref{tab:tab2}). 

Run~4 (where $Q_{\rm A}=3$) and run~7 (where $Q_{\rm B}=12.5$) do not satisfy the first criterion (because of the large values of $\sigma{}_{\rm A}$ and $\sigma{}_{\rm B}$, respectively). While $P_{\rm KS}$ is $<0.05$ in runs~11 ($\Delta{}v=3.5$ km s$^{-1}$) and 16 ($f_{\rm bin}=0.75$). 
From Figure~\ref{fig:fig2} and Table~\ref{tab:tab2}, it is apparent that some of our models match the kinematics of the Gamma Velorum cluster quite well. Furthermore, we are able to put constraints on the most relevant initial parameters. 
 In particular, if the initial velocity shift is $\Delta{}v>3$ km s$^{-1}$ (runs~10 and 11), then $\Delta{}$RV is larger than the observed one by more than 1 $\sigma{}$, for the entire simulation. The tidal forces exerted during the interaction between pop.~A and pop.~B are not sufficient to quench this velocity significantly. If $\Delta{}v\ge{}3.5$ km s$^{-1}$, $P_{\rm KS}$  is very low ($<0.05$). 

  More interestingly, if the simulated pop.~B is close to virial equilibrium ($Q_{\rm B}\le{}2$, runs~5 and 6), then its velocity dispersion ($\le{}0.7$ km s$^{-1}$) is much smaller than the observed one ($1.60\pm{}0.37$ km s$^{-1}$), and the RV distributions of the two populations appear almost completely separated (Figure~\ref{fig:fig2}). On the other hand, if $Q_{\rm B}\ge{}12.5$, then the simulated velocity dispersion of pop.~B is too large with respect to the observed one ($\sigma{}_{\rm B}\sim{}2.8$  km s$^{-1}$ for $Q_{\rm B}=12.5$ in run~7, i.e. $3\,{}\sigma{}$ higher than the observed value). Thus, the model can reproduce the observed RV distribution only if pop.~B is significantly hotter than a virial system, but not too hot ($2<Q_{\rm B}<12.5$). This implies that pop.~B is unbound, likely as a consequence of the evaporation of the parent molecular gas. 

Other constraints can be put on the virial ratio of pop.~A: if $M_{\rm A}\sim{}500$ M$_\odot$ and $Q_{\rm A}\ge{}3.0$, then the velocity dispersion of the simulated pop.~A ($\sim{}1$ km s$^{-1}$, run~4) is too large to be consistent with the observed one ($0.34\pm{}0.16$ km s$^{-1}$). In contrast, values of $Q_{\rm A}<2.0$ are consistent with the data, indicating that pop.~A is approximately in virial equilibrium. 

The binary fraction $f_{\rm bin}$ is very important for the comparison with observations. In fact, unresolved binaries tend to broaden the RV distribution (we recall that in Figure~\ref{fig:fig2} we plot even the simulated binaries as unresolved, for a better comparison with the data). In most simulations, we assume $f_{\rm bin}=0.46$ for analogy with the results of \cite{Jeffries2014}, but we run two test cases with $f_{\rm bin}=0$ (run~15) and 0.75  (run~16), respectively. The effect of $f_{\rm bin}$  is particularly important for pop.~A. If $f_{\rm bin}=0$ (run~15), then the scatter in the RVs of pop.~A is much smaller: a larger value of $Q_{\rm A}$ is requested to reproduce $\sigma{}_{\rm A}$ if $f_{\rm bin}=0$ (run~15), but the simulations still fail to reproduce the high-velocity tail of the RV distribution. In contrast, if $f_{\rm bin}=0.75$ (run~16), the RV distribution is broader, and a lower value of $Q_{\rm A}$ is requested to match the data.

The characteristic radii of the two populations are more important for the spatial distribution than for the RV. On the other hand, the characteristic radii have also an impact on the RV, and there
 is partial degeneracy with the effect of the virial ratio. For example, if pop.~B is initially spread over a larger radius ($r_{\rm B}=2$ pc, run~18), for the same initial mass $M_{\rm B}$, then its velocity dispersion will be lower. Thus, the observations can be matched only for a larger value of $Q_{\rm B}$, but this affects the disruption timescale of pop.~B.

Finally, the mass of the two populations is another important quantity, and there is partial degeneracy with the virial ratio. For example, run~12 has the same parameters as run~1, but the mass of the two clusters is lower by a factor of $\sim{}5$. The RV distributions of pop~A. and pop.~B partially overlap in run~1, while they are well separated in run~12. The result is that  run~12 does not match the data (especially $\sigma{}_{\rm B}$ and $\Delta{}$RV, Table~\ref{tab:tab2}). In case of a lower initial total mass, larger values of $Q_{\rm A}$ and   $Q_{\rm B}$ are requested, to obtain a best matching with the data. For example, in our simulations, if $M_{\rm A}+M_{\rm B}\sim{}150$ M$_\odot{}$ instead of $\sim{}800$ M$_\odot$, we need  $Q_{\rm A}\ge{}4$ and   $Q_{\rm B}\ge{}8$ (as in run~13, Figure~\ref{fig:fig2}) to reproduce the Gamma Velorum data. 

The values adopted in run~13 match  the current mass of the Gamma Velorum cluster quite well, but present an issue: they can hardly account for the formation of the $\gamma{}^2$ Velorum WR-O binary. In Table~\ref{tab:tab1}, we report the masses of the members of the most massive binary in each run. The most massive binaries in runs~12 and 13 are well below 10 M$_\odot$. Even if we account for stochastic fluctuations, the formation of a $\sim{}(30,\,{}30)$ M$_\odot$ binary in clusters with initial mass $\sim{}100$ M$_\odot$ is very unlikely. Thus, we can argue either that  the formation of $\gamma{}^2$ Velorum is not related to the rest of the Gamma Velorum cluster (which is quite unlikely), or that  the initial mass of the clusters was a factor of $\gtrsim{}5$ higher, and then the partial cluster dissolution due to gas evaporation has led to a smaller total mass. Our runs with virial ratio $>0.5$ account for this effect of gas evaporation and ongoing cluster dissolution. Furthermore, our simulated YSCs evolve dynamically and progressively spread over an area larger than the one observed by the GES. This partially explains the difference between the mass estimated by the observations (within the 0.9 deg$^2$ area surveyed by GES) and the total mass of the simulated YSCs in run~1 (see the discussion in Sec.~\ref{sec:shift}).




\begin{figure}
\center{{
\epsfig{figure=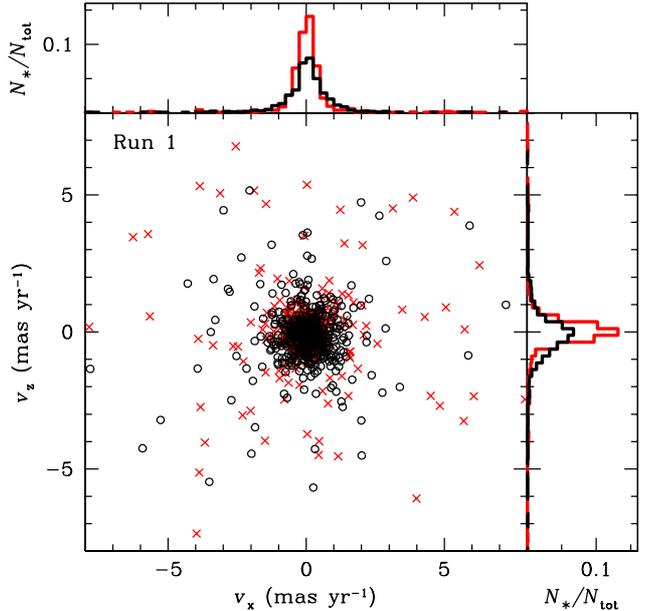,width=9cm} 
}}
\caption{\label{fig:fig3}
Proper motions in the plane of the sky for run~1 at $t=4.8$ Myr. In the main panel, red crosses: pop.~A, black open circles: pop.~B. In the marginal histograms, red (black) histogram: pop.~A (pop.~B). On the $y$ axis of the marginal histograms: number of stars per bin, normalized to the total number of stars in each cluster ($N_\ast{}/N_{\rm tot}$). The simulated proper motions shown in this Figure do not include observational uncertainties.}
\end{figure}
 Figure~\ref{fig:fig3} shows the proper motions in the plane of the sky (the $xz$ plane, according to our convention) for run~1.  Pop.~A and pop.~B are not significantly decoupled, if we consider the proper motions. On the other hand, since pop.~B is more supervirial than pop.~A, the distribution of proper motions has a larger dispersion for pop.~B ($\tilde{\sigma}_{\rm x,\,{}B}=0.96$ km s$^{-1}$, $\tilde{\sigma}_{\rm z,\,{}B}=1.00$ km s$^{-1}$, where $\tilde{\sigma}_{\rm x,\,{}B}$ and $\tilde{\sigma}_{\rm z,\,{}B}$ are the intrinsic velocity dispersions of pop.~B along the $x$ and the $z$ axis, respectively)  than for pop.~A ($\tilde{\sigma}_{\rm x,\,{}A}=0.46$ km s$^{-1}$, $\tilde{\sigma}_{\rm z,\,{}A}=0.47$ km s$^{-1}$, where $\tilde{\sigma}_{\rm x,\,{}A}$ and $\tilde{\sigma}_{\rm z,\,{}A}$ are the intrinsic velocity dispersions of pop.~A along the $x$ and the $z$ axis, respectively). Furthermore, if the plane of the sky is slightly different from the assumed one (i.e. if the relative velocity vector between the two YSCs is not perfectly aligned to the line-of-sight), we expect to observe a further difference between pop.~A and pop.~B. Gaia data will be crucial to test this feature.

\subsection{Spatial distribution}\label{sec:shift}
\begin{figure}
\center{{
\epsfig{figure=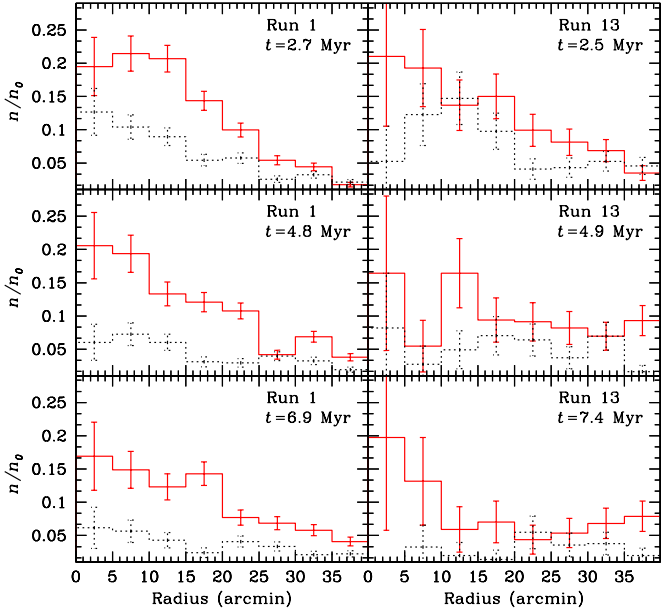,width=9cm} 
}}
\caption{\label{fig:fig4}
Normalized radial stellar density  in run~1 (left-hand panels) and in run~13 (right-hand panels) at different times. The simulated cluster is assumed to be at a distance of 350 pc. Solid red line: pop.~A; dotted black line: pop.~B. The density $n$ is calculated as stars per arcmin$^2$, within concentric annuli. The normalization $n_0$ was chosen so that the total area below the histogram  of pop.~A is one. The error bars account for Poisson uncertainties. 
}
\end{figure}
In the data published by \cite{Jeffries2014}, pop.~A appears to be marginally more concentrated than pop.~B, and the two centroids are offset by $\sim{}4.4\pm{}3.0$ arcmin ($\sim{}0.45\pm{}0.31$ pc). Our simulations can naturally explain the radial offset and the different centroid of the two populations of the Gamma Velorum cluster: in our models, pop.~A and pop.~B come from two different sub-clusters, which might be born with different concentration and are expected to have different centroids.

 Figure~\ref{fig:fig4} shows the normalized radial density of stars in pop.~A and pop.~B in runs~1 and 13, at different times. Pop.~B is slightly less concentrated than pop.~A.  Furthermore, the concentration of pop.~B diminishes with time, as a result of the fact that the second (less massive) cluster is being tidally disrupted by the first (more massive) cluster (see also Figure~\ref{fig:fig1}).  The probability that the chance deviation between the spatial distribution of pop.~A and that of pop.~B is expected to be larger, according to a Kolmogorov-Smirnov test, is $P_{\rm KS}<10^{-5}$ and $<0.1$  for run~1 and run~13, respectively. Thus, the two simulated populations follow different spatial distributions. In the other runs, we observe the same trend, even if the relative difference between  pop.~A and pop.~B changes (depending on the initial conditions). The simulated spatial distributions (Figure~\ref{fig:fig4}) are qualitatively similar to the observed radial spatial distribution of the Gamma Velorum cluster, as shown in Figure~11 of \cite{Jeffries2014}, even if we recall that these two figures cannot be directly compared\footnote{Figure~11 of \cite{Jeffries2014} and our Figure~\ref{fig:fig4} cannot be directly compared, since we know the intrinsic membership of each single star to pop.~A (i.e. cluster~1) or pop.~B (i.e. cluster~2), while the selection by \cite{Jeffries2014} is based on a probabilistic approach. As we have discussed in Section~\ref{sec:RV}, a  fraction of genuine members of our simulated pop.~A would be attributed to pop.~B, based on the statistical approach by \cite{Jeffries2014}.}.

Column~11 of Table~\ref{tab:tab2} shows that there is an offset of $0-1.5$ pc between the centres-of-mass of the two simulated clusters in the $xz$ plane (i.e. in the assumed plane of the sky) at $0\le{}t/{\rm Myr}\le{}8$, depending on the initial conditions. 
For most runs, the offset is consistent with the observed one ($\sim{}0.45\pm{}0.31$ pc). 
 
Finally, Table~\ref{tab:tab2} also shows that there is an offset of $0-18$ pc between the centres-of-mass of the two simulated clusters along the $y$ axis (i.e. along the line-of-sight) at $0\le{}t/{\rm Myr}\le{}8$. 
The colour-magnitude diagram of the Gamma Velorum cluster is consistent with an offset of $4\pm{}5$ pc (along the line-of-sight), if we assume that pop.~A and pop.~B are coeval (or with an even larger offset, if we assume that pop.~A is slightly older, as indicated by the Li depletion, \citealt{Jeffries2014}). Thus, the simulated offset along the line-of-sight is fairly consistent with the one indicated by the observations.
\begin{figure*}
\center{{
\epsfig{figure=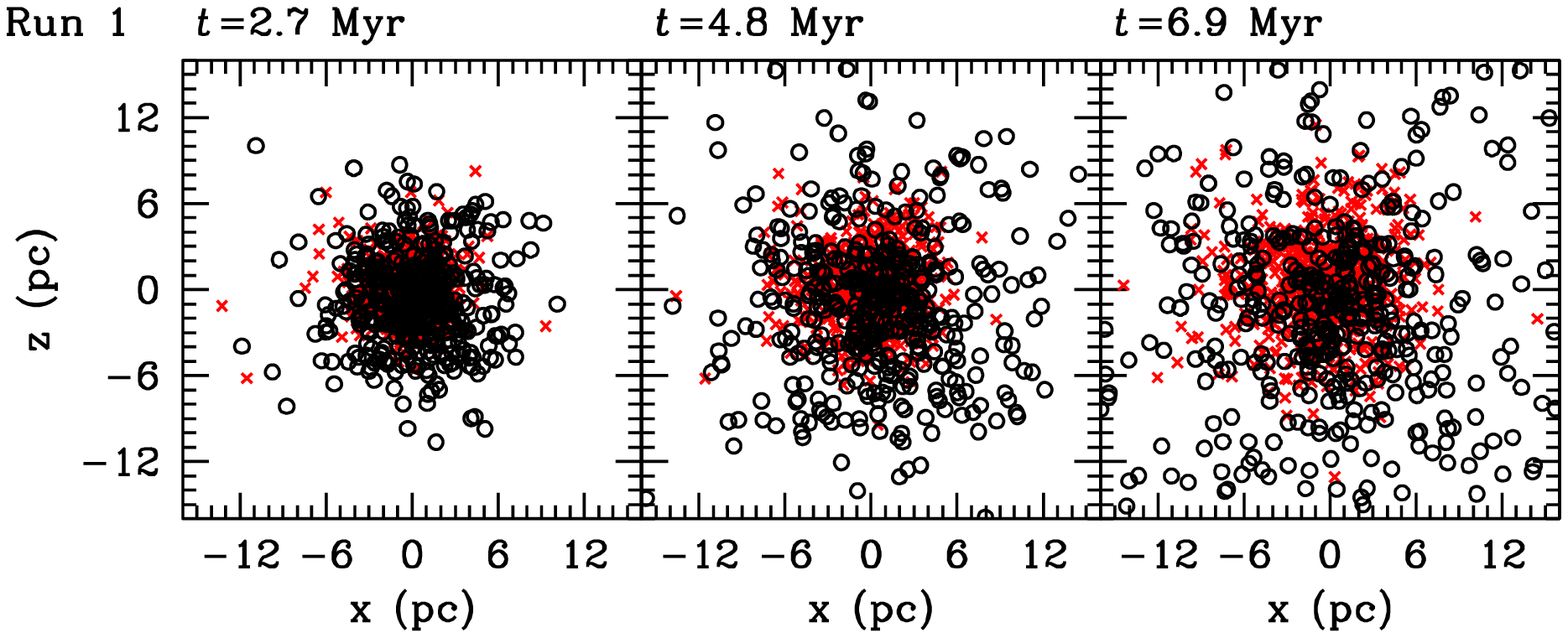,width=18cm} 
}}
\caption{\label{fig:fig5}
Position of simulated stars belonging to pop.~A (red crosses) and pop.~B (black open circles), projected in the plane of the sky (the $xz$ plane of the simulation), at three different times ($t=2.5,\,{}4.8$ and 6.9 Myr since the beginning of the simulation) in run~1. The centre of the frame is the centre of mass of the entire cluster (pop.~A+ pop.~B). We note that pop.~B expands much faster than pop.~A given its supervirial condition. Stars belonging to pop.~B can be found at $>10$ pc from the centre of pop.~A.
}
\end{figure*}

\begin{table*}
\begin{scriptsize}
\begin{center}
\caption{Main results of the $N-$body simulations.}
 \leavevmode
\begin{tabular}{lllllllllllll}
 \hline 
Run   & $t$  & $\sigma{}_{\rm A}$  & $\sigma{}_{\rm B}$  & $\Delta{}$RV  & $f_{\rm A}$ & $\tilde{\sigma{}}_{\rm A}$  & $\tilde{\sigma{}}_{\rm B}$  &  $\tilde{\Delta{}}$RV & $\Delta{}y$ & $\Delta{}xz$ & $P_{\rm KS}$ \\
  &  (Myr) & (km s$^{-1}$) & (km s$^{-1}$) &  (km s$^{-1}$) & &  (km s$^{-1}$) &  (km s$^{-1}$) &  (km s$^{-1}$) &  (pc) & (pc) & \\
\hline
Run~1 & 2.7   &  0.46 & 1.93 & 1.95 & 0.35 & 0.54 & 1.29 & 2.48 & 1.77  & 0.01 & 0.72\\
Run~1 & 4.8   &  0.43 & 1.84 & 1.93 & 0.37 & 0.50 & 1.25 & 2.42 & 7.22  & 0.16 & 0.84\\
Run~1 & 6.9   &  0.38 & 1.85 & 1.82 & 0.35 & 0.49 & 1.23 & 2.41 & 12.37 & 0.55 & 0.84
\vspace{0.1cm}\\
Run~2 & 2.7   &  0.49 & 1.21 & 2.66 & 0.55 & 0.57 & 1.37 & 2.51 & 1.41 & 0.45 & 0.86 \\
Run~2 & 4.8   &  0.42 & 1.66 & 2.00 & 0.40 & 0.53 & 1.33 & 2.45 & 6.90 & 0.12 & 0.84 \\
Run~2 & 6.9   &  0.47 & 1.68 & 2.02 & 0.42 & 0.50 & 1.32 & 2.42 & 12.00 & 0.06 & 0.84 \vspace{0.1cm}\\

Run~3 & 2.6   &  0.57 & 1.98 & 1.63 & 0.32 & 0.66 & 1.23 & 2.46 & 1.92 & 0.12 & 0.53\\
Run~3 & 4.8   &  0.50 & 1.90 & 1.67 & 0.31 & 0.61 & 1.21 & 2.41 & 6.97 & 0.58 & 0.56 \\
Run~3 & 6.9   &  0.51 & 1.85 & 1.67 & 0.33 & 0.59 & 1.20 & 2.39 & 12.09 & 0.82 & 0.68\vspace{0.1cm}\\

Run~4 & 2.7   &  1.04 ($\ast{}$) & 1.38 & 2.66 & 0.53 & 0.95 & 1.30 & 2.55 & 1.71 & 0.06 & 0.09\\
Run~4 & 4.8   &  0.95 ($\ast{}$) & 1.33 & 2.58 & 0.51 & 0.92 & 1.27 & 2.51 & 6.94 & 0.06 & 0.11\\
Run~4 & 6.9   &  0.94 ($\ast{}$) & 1.30 & 2.57 & 0.51 & 0.91 & 1.26 & 2.50 & 12.51 & 0.46 & 0.12\vspace{0.1cm}\\

Run~5 & 2.7   &  0.56 & 0.55 & 2.70 & 0.58 & 0.52 & 0.73 & 2.58 & 1.91 & 0.04 & 0.24\\
Run~5 & 4.8   &  0.51 & 0.59 & 2.32 & 0.55 & 0.51 & 0.69 & 2.24 & 7.08 & 0.25 & 0.54\\ 
Run~5 & 6.9   &  0.50 & 0.51 & 2.22 & 0.56 & 0.49 & 0.68 & 2.15 & 11.54 & 1.36 & 0.49\vspace{0.1cm}\\

Run~6 & 2.7   &  0.58 & 0.69 & 2.74 & 0.56 & 0.54 & 0.73 & 2.65 & 1.80 & 0.05 & 0.27\\
Run~6 & 4.8   &  0.52 & 0.67 & 2.57 & 0.56 & 0.50 & 0.69 & 2.50 & 7.35 & 0.10 & 0.37\\ 
Run~6 & 6.9   &  0.51 & 0.61 & 2.51 & 0.56 & 0.55 & 0.69 & 2.46 & 12.60 & 0.16 & 0.40\vspace{0.1cm}\\


Run~7 & 2.7   &  0.53 &  2.79 ($\ast{}$) & 2.21 & 0.42 & 0.55 & 2.20 & 2.80 & 1.62 & 0.03 & 0.06\\
Run~7 & 4.8   &  0.47 &  2.78 ($\ast{}$) & 2.21 & 0.43 & 0.50 & 2.18 & 2.80 & 7.01 & 0.10 & 0.09\\
Run~7 & 6.9   &  0.44 &  2.79 ($\ast{}$) & 2.11 & 0.42 & 0.53 & 2.18 & 2.78 & 11.43 & 0.72 & 0.13\vspace{0.1cm}\\

Run~8 & 2.7   &  0.49 & 1.65 & 1.25 & 0.36 & 0.55 & 1.27 & 1.67 & -0.50 & 0.01 & 0.72\\
Run~8 & 4.8   &  0.47 & 1.62 & 1.20 & 0.38 & 0.50 & 1.21 & 1.64 & 3.27 & 0.19 & 0.70 \\
Run~8 & 6.9   &  0.43 & 1.57 & 1.18 & 0.38 & 0.46 & 1.19 & 1.61 & 6.39 & 0.5 & 0.40 \vspace{0.1cm}\\

Run~9 & 2.7   &  0.46 & 1.80 & 1.62 & 0.34 & 0.55 & 1.28 & 2.12 & 0.77 & 0.01 & 0.79\\
Run~9 & 4.8   &  0.44 & 1.70 & 1.62 & 0.36 & 0.50 & 1.22 & 2.07 & 5.48 & 0.16 & 0.76\\
Run~9 & 6.9   &  0.41 & 1.72 & 1.53 & 0.37 & 0.46 & 1.21 & 2.04 & 9.57 & 0.53 & 0.84 \vspace{0.1cm}\\

Run~10 & 2.7   &  0.58 & 1.35 & 3.04 & 0.52 & 0.54 & 1.33 & 3.01 & 3.23 & 0.03 & 0.23 \\
Run~10 & 4.8   &  0.55 & 1.27 & 2.99 & 0.54 & 0.50 & 1.30 & 2.95 & 9.77 & 0.15 & 0.35 \\
Run~10 & 6.9   &  0.49 & 1.26 & 2.93 & 0.52 & 0.48 & 1.28 & 2.93 & 15.58 & 0.50 & 0.33 \vspace{0.1cm}\\

Run~11 & 2.7   &  0.57 & 1.37 & 3.49 & 0.52 & 0.55 & 1.39 & 3.44 & 4.46 & 0.02 & 0.02 \\
Run~11 & 4.8   &  0.53 & 1.37 & 3.41 & 0.52 & 0.50 & 1.36 & 3.39 & 11.96 & 0.13 & 0.04  \\ 
Run~11 & 6.9   &  0.48 & 1.35 & 3.35 & 0.52 & 0.52 & 1.34 & 3.37 & 18.44 & 0.70 & 0.04  \vspace{0.1cm}\\

Run~12 & 2.4   &  0.31 & 0.64 & 2.40 & 0.54 & 0.29 & 0.69 & 2.27 & 0.87 & 0.21 & 0.72\\ 
Run~12 & 4.8   &  0.23 & 0.64 & 2.40 & 0.53 & 0.24 & 0.66 & 2.25 & 6.48 & 0.34 & 0.77\\ 
Run~12 & 7.2   &  0.26 & 0.72 & 2.36 & 0.54 & 0.24 & 0.70 & 2.22 & 11.96 & 0.53 & 0.81\vspace{0.1cm}\\ 

Run~13 & 2.5   &  0.33 & 1.69 & 1.86 & 0.37 & 0.48 & 0.88 & 2.30 & 0.64 & 0.01 & 0.88 \\ 
Run~13 & 4.9   &  0.37 & 1.61 & 1.95 & 0.39 & 0.46 & 0.87 & 2.27 & 6.31 & 0.02 & 0.92 \\
Run~13 & 7.4   &  0.48 & 0.98 & 2.35 & 0.56 & 0.45 & 0.86 & 2.26 & 11.94 & 0.03 & 0.89 \vspace{0.1cm}\\ 

Run~14 & 2.7   &  0.48 & 1.95 & 1.79 & 0.35 & 0.56 & 1.35 & 2.42 & -3.13 & 0.02 & 0.75\\
Run~14 & 4.8   &  0.46 & 1.83 & 2.04 & 0.40 & 0.50 & 1.30 & 2.42 & 2.30 & 0.19 & 0.78\\
Run~14 & 6.9   &  0.42 & 1.80 & 1.96 & 0.39 & 0.48 & 1.28 & 2.40 & 7.02 & 0.45 & 0.81\vspace{0.1cm}\\

Run~15 & 2.6   & 0.53 & 1.07 & 2.83 & 0.59 & 0.55 & 1.38 & 2.44 & 1.86 & 0.01 & 0.42\\
Run~15 &  4.7   &  0.40 & 1.23 & 2.43 & 0.56 & 0.46 & 1.34 & 2.38 & 7.37 & 0.01 & 0.25\\
Run~15 &  6.8   &  0.35 & 1.28 & 2.34 & 0.54 & 0.43 & 1.33 & 2.36 & 12.78 & 0.02 & 0.24\vspace{0.1cm}\\

Run~16 & 2.7   &  0.61 & 1.29 & 2.87 & 0.50 & 0.50 & 1.20 & 2.75 & 1.94 & 0.07 & 0.03 \\
Run~16 & 4.8   &  0.39  & 2.19 & 2.01 & 0.28 & 0.44 & 1.17 & 2.68 & 6.11 & 0.41 & 0.03 \\
Run~16 & 6.9   &  0.37 & 2.13 & 1.86 & 0.28 & 0.42 & 1.16 & 2.61 & 10.50 & 0.44 & 0.05\vspace{0.1cm}\\

Run~17 & 2.7   &  0.64 & 1.20 & 2.61 & 0.57 & 0.56 & 1.35 & 2.42 & 2.12 & 0.07 & 0.64\\

Run~17 & 4.8   &  0.37 & 1.84 & 1.83 & 0.35 & 0.50 & 1.30 & 2.42 & 7.46 & 0.24 & 0.67\\
Run~17 & 7.0   &  0.35 & 1.75 & 1.96 & 0.38 & 0.48 & 1.28 & 2.40 & 12.63 &  0.29 & 0.67\vspace{0.1cm}\\

Run~18 & 2.7   &  0.57 & 0.92 & 2.51 & 0.54 & 0.54 & 0.97 & 2.36 & 1.71 & 0.02 & 0.51\\
Run~18 & 4.8   &  0.50 & 0.92 & 2.36 & 0.53 & 0.49 & 0.90 & 2.31 & 7.11 & 0.17 & 0.57\\
Run~18 & 6.9   &  0.48 & 0.93 & 2.33 & 0.53 & 0.46 & 0.90 & 2.26  & 11.84 & 0.51 & 0.78\vspace{0.1cm}\\

Run~19 & 2.4   &  0.51 & 2.11 & 1.62 & 0.39 & 0.59  & 1.58  & 1.72 & 1.77 & 0.01 & 0.76\\ 
Run~19 & 4.8   &  0.44 & 1.95 & 1.71 & 0.42 & 0.49  & 1.53  & 1.69 & 8.50 & 0.17 & 0.86\\
Run~19 & 6.8   &  0.39 & 1.84 & 1.79 & 0.44 & 0.44  & 1.51  & 1.67 & 13.70 & 0.80 & 0.82\vspace{0.1cm}\\ 

Run~20 & 2.5   &  0.48 & 1.83 & 1.81 & 0.30 & 0.56 & 1.34 & 2.20 & 1.44 & 0.03 & 0.45\\
Run~20 & 5.0   &  0.44 & 1.72 & 1.83 & 0.33 & 0.49 & 1.29 & 2.17 & 8.03 & 0.12 & 0.60\\
Run~20 & 7.1   &  0.39 & 1.82 & 1.67 & 0.29 & 0.49 & 1.29 & 2.14 & 12.44 & 1.00 & 0.71\vspace{0.1cm}\\

Run~21 & 2.5   &  0.49 & 1.84 & 1.53 & 0.34 & 0.56 & 1.33 & 2.13 & 10.40 & 0.02 & 0.84\\
Run~21 & 5.0   &  0.49 & 1.82 & 1.62 & 0.38 & 0.52 & 1.33 & 2.07 & 15.10 & 0.19 & 0.81\\
Run~21 & 7.1   &  0.41 & 1.84 & 1.46 & 0.35 & 0.49 & 1.33 & 2.06 &  19.37 ($\ast{}$) & 0.49 & 0.84 \vspace{0.1cm}\\

Data$^{\rm a}$   & -- & $0.34\pm{}0.16$ &  $1.60\pm{}0.37$ & $2.15\pm{}0.48$ & $0.48\pm{}0.11$ & --  & -- & -- & $4\pm{}5$$^{\rm b}$ & $0.45\pm{}0.31$ & -- \vspace{0.1cm} \\
\noalign{\vspace{0.1cm}}
\hline
\end{tabular}
\begin{flushleft}
\footnotesize{Run (column 1): identifying name of the run;  $t$ (column 2): time elapsed since the beginning of the simulation; $\sigma{}_{\rm A}$ (column 3): velocity dispersion of pop.~A as derived from fitting with two Gaussian components;  $\sigma{}_{\rm B}$(column 4): velocity dispersion of pop.~B as derived from fitting with two Gaussian components; $\Delta{}$RV (column 5): RV difference between pop.~A and pop.~B as derived from fitting with two Gaussian components; $f_{\rm A}$ (column 6): fraction of stars belonging to pop.~A, as derived from fitting with two Gaussian components; $\tilde{\sigma{}}_{\rm A}$ (column 7): intrinsic velocity dispersion of pop.~A; $\tilde{\sigma{}}_{\rm B}$  (column 8): intrinsic velocity dispersion of pop.~B;  $\tilde{\Delta{}}$RV (column 9): difference between the average RV of pop.~A and pop.~B; $\Delta{}y$ (column 10): distance between the centre of mass of pop.~A and the centre-of-mass of pop.~B along the line-of-sight ($y$ axis). A negative value of $\Delta{}y$ means that the first collision between cluster 1 and cluster 2 has not happened yet; $\Delta{}xz$ (column 11): distance between the centre of mass of pop.~A and the centre-of-mass of pop.~B in the plane of the sky ($xz$ plane); $P_{\rm KS}$ (column 12): probability that the chance deviation between the observed RV distribution and the simulated RV distribution is expected to be larger, according to a Kolmogorov-Smirnov test. $P_{\rm KS}$ is very small ($<0.05$) only for run~11 ($\Delta{}v=3.5$ km s$^{-1}$) and run~16 ($f_{\rm bin}=0.75$).\\
$^{\rm a}$ The values reported in the last line come from the analysis of \cite{Jeffries2014}. $^{\rm b}$The observed pop.~A in Gamma Velorum is closer by $\Delta{}y=4\pm{}5$ pc than pop.~B, if the observed difference in the colour-magnitude diagram (\citealt{Jeffries2014}) is due to a distance difference rather than to an age difference. \\ Quantities differing by $>3\,{}\sigma{}$ with respect to the observed value are marked by  ($\ast{}$).\label{tab:tab2}}
\end{flushleft}
\end{center}
\end{scriptsize}
\end{table*}
 Figure~\ref{fig:fig5} shows the position of simulated stars belonging to pop.~A (red crosses) and pop.~B (black circles), projected in the plane of the sky (the $xz$ plane of the simulation), at three different times ($t=2.5,\,{}4.8$ and 6.9 Myr since the beginning of the simulation) in run~1. It is apparent that pop.~B expands faster than pop.~A, because of its supervirial condition. We note that stars of pop.~B can be found at $>10$ pc from the centre of pop.~A.   Recently,  \citet{Sacco2014} found that 15 stars, in the direction of the NGC~2547 cluster, i.e. about two degrees ($\sim{}10$ pc) south of $\gamma{}^2$ Velorum, have the same properties as the pop.~B members of the Gamma Velorum cluster. To compare our simulations with the results of \citet{Sacco2014}, we select a box in our run~1 (at $t=4.8$ Myr) with approximately the same size and the same location (with respect to the centre of the cluster) as the area analyzed by \citet{Sacco2014}. We count $16$ pop.~B objects (6 of which are binaries) in this box. This is in excellent agreement with the result of \citet{Sacco2014}, who find $15$ pop.~B members in the field of NGC~2547. This result supports our scenario, and strengthens the evidence that pop.~B is strongly supervirial. Figure~\ref{fig:fig5} also shows that our simulated star clusters are more extended than the area reported by Jeffries et al. (2014, $\sim{}0.9$ deg$^2$). It is reasonable to expect that the outer rim of the Gamma Velorum cluster is much larger than the area observed by \citet{Jeffries2014}.

\subsection{Collision or line-of-sight superposition?}
Our models are able to reproduce most properties of the Gamma Velorum cluster, provided that pop.~B is significantly supervirial. However, we may wonder whether we really need a physical collision between the two sub-clusters. Is it possible to explain the properties of the Gamma Velorum cluster with a mere line-of-sight superposition between pop.~A and pop.~B, without requiring any interactions between the two sub-clusters? 

To answer this question, we have performed run~21, in which the two sub-clusters form close to each other (5 pc) but increase their separation rather than collide. The initial properties of run~21 are the same as those of run~1 (which matches the properties of the Gamma Velorum cluster particularly well), apart from the sign of the relative velocity. The main kinematic properties of run~21 (i.e. the values of $\sigma{}_{\rm A}$, $\sigma{}_{\rm B}$, and $\Delta{}$RV) are very similar to those of run~1, indicating that the velocity field of the two populations is not severely affected by the collision. The main reason is that the relative velocity of the two sub-clusters is large with respect to their internal kinematics. On the other hand, the line-of-sight distance between the two sub-clusters becomes too large ($>19$ pc, $3\,{}\sigma{}$ larger than the observed value) at $t\gtrsim{}7.1$ Myr. 

The same conclusion can be reached on the basis of a back-of-the-envelope calculation. The GES data indicate that there is an offset $\Delta{}RV=2.15\pm{}0.48$ km s$^{-1}$ between the RV of pop.~A and pop.~B. Furthermore, pop.~A might be closer to us by $4\pm{}5$ pc. If we assume that pop.~A and pop.~B formed in the same place and then drifted away from each other, pop.~B is now at a distance $d_{\rm AB}$ from pop.~A:
\begin{equation}
d_{\rm AB}=10\,{}{\rm pc}\,{}\left(\frac{\Delta{}RV}{2\,{}{\rm km}\,{}{\rm s}^{-1}}\right)\,{}\left(\frac{t_{\rm B}}{5\,{}{\rm Myr}}\right),
\end{equation}
where $t_{\rm B}$ is the age of pop.~B. Thus, even if pop.~A and pop.~B formed in the same place 5 Myr ago, $d_{\rm AB}$ is already $1\,{}\sigma{}$ larger than the observed displacement between pop.~A and pop.~B. If pop.~B formed farther from us than pop.~A, then we would expect a larger  observed displacement between pop.~A and pop.~B.
On the other hand, if pop.~A and pop.~B formed in the same place, they should have interacted between each other before drifting away. For the same reason, if pop.~B formed closer to us, it should have interacted with pop.~A before drifting away. This favours the scenario of a gravitational encounter between the two sub-clusters. This argument is weakened by the fact that our measurement of the line-of-sight displacement between  pop.~A and pop.~B relies on the colour-magnitude diagram and is quite inaccurate. Parallax measurements by Gaia will give invaluable hints on this point.

\begin{figure}
\center{{
\epsfig{figure=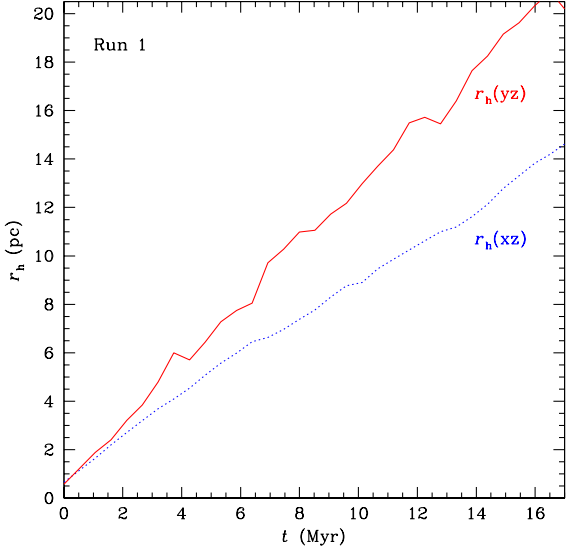,width=8cm} 
}}
\caption{\label{fig:figtid}
Half-mass radius of pop.~B as a function of time in run~1. $r_{\rm h}(xz)$: half-mass radius projected in the $xz$ plane. $r_{\rm h}(yz)$: half-mass radius projected in the $yz$ plane (where the tidal force exerted by pop.~A is maximum). 
}
\end{figure}

Do the simulations show any other quantitative difference between a simple line-of-sight superposition and  a gravitational interaction between pop.~A and pop.~B? A physical collision leaves an imprint on the structure of the two sub-clusters, and especially of the lighter one. In our simulations, pop.~B is tidally perturbed and stretched by the gravitational encounter with the more massive pop.~A. The tidal perturbation can be quantified by comparing the half-mass radius projected in the plane where the tidal force is maximum ($r_{\rm h}(yz)$ or $r_{\rm h}(xy)$), with the half-mass radius projected in the plane where the tidal force is minimum ($r_{\rm h}(xz)$). Figure~\ref{fig:figtid} shows that $r_{\rm h}(yz)$ is significantly larger than $r_{\rm hm}(xz)$, indicating that pop.~B is severely stretched by the interaction. Again, this might be checked with forthcoming parallax measurements  by Gaia.

In summary, a collision scenario and a mere line-of-sight superposition show no significant differences in the velocity dispersions and in the RV distribution. The only way to distinguish between these two scenarios is to measure the line-of-sight displacement between the two populations and any possible tidal deformation of pop.~B. 

\subsection{Relaxing the assumption of line-of-sight superposition}\label{sec:LOS}
\begin{figure}
\center{{
\epsfig{figure=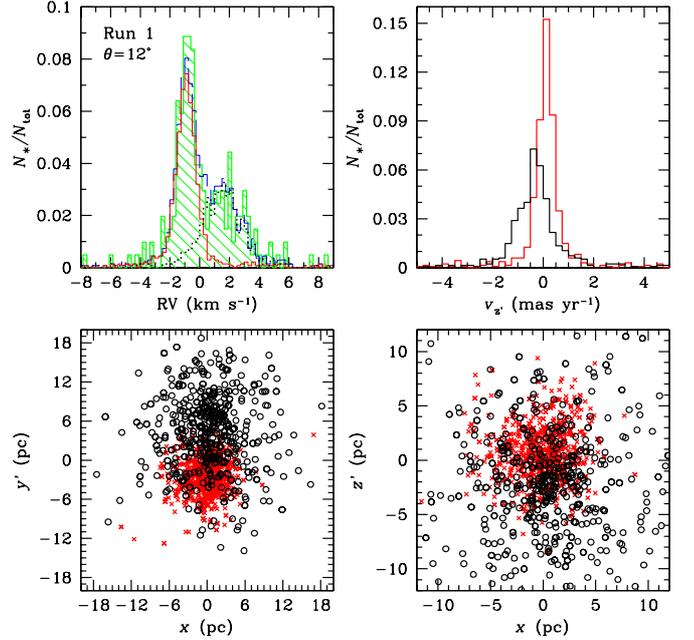,width=9cm} 
}}
\caption{\label{fig:fig7}
Properties of run~1 (at $t=4.8$ Myr) if rotated by an angle  $\theta{}=12^\circ{}$ about the $x$ axis. Top left-hand panel: RV distribution (symbols are the same as in Fig.~\ref{fig:fig2}); top right-hand panel: distribution of proper motions along $z'$ (symbols are the same as in the marginal histograms of Fig.~\ref{fig:fig3}, in particular pop.~A and pop.~B correspond to the red and black histogram, respectively);  bottom left-hand panel: positions in the $xy'$ plane (symbols are the same as in Fig.~\ref{fig:fig1}, in particular pop.~A and pop.~B correspond to red crosses and black open circles, respectively);  bottom right-hand panel: positions in the $xz'$ plane (symbols are the same as in Fig.~\ref{fig:fig5}, in particular pop.~A and pop.~B correspond to red crosses and black open circles, respectively).
}
\end{figure}

 It is very unlikely that the two sub-structures of a YSC are observed exactly along the line-of-sight. How much can we relax the assumption of line-of-sight superimposition and still match the observed data? To check this hypothesis, we focus on our fiducial run (run~1), and change the line-of-sight by rotating the $yz$ plane about the $x-$axis by an angle $\theta{}$. We stop rotating when at least one of our reference quantities for the comparison with the data ($\sigma{}_{\rm A}$, $\sigma{}_{\rm B}$, $\Delta{}$RV, $f_{\rm A}$, $\Delta{}y$ and  $\Delta{}xz$) differs from the best observed value by more than 3 $\sigma{}$. The first parameter for which the discrepancy becomes $>3\,{}\sigma{}$ is the difference between the centroids ($\Delta{}xz$). At an angle $\theta{}=12^\circ$ between the original $y$ axis and the new $y$ axis ($y'$), we obtain  $\sigma{}_{\rm A}=0.48$ km s$^{-1}$, $\sigma{}_{\rm B}=1.64$ km s$^{-1}$, $\Delta{}$RV=1.99 km s$^{-1}$, $f_{\rm A}=0.40$, $\Delta{}y'=7.07$ pc and  $\Delta{}xz'=1.50$ pc (the best value for the observations is $\Delta{}xz=0.45\pm{}0.31$ pc, see Table~2). Thus, we can conclude that configurations in which the two sub-clusters are superimposed or slightly offset (by an angle $\theta{}\le{}12^\circ{}$) reasonably match the data. Larger angles are possible, but only for different collision times. Figure~\ref{fig:fig7} shows the RV distribution (along $y'$), the distribution of proper motions along ${\rm z'}$, and the projected positions (in the $xy'$ and $xz'$ planes) if run~1 is rotated by $\theta{}=12^\circ{}$ about the $x$ axis.

The probability of observing two sub-clusters offset by an angle $\theta{}\le{}12^\circ{}$ is $\sim{}0.022$. 
On the other hand, we recall that while the centroid of pop.~A is quite well constrained, the centroid of pop.~B is more uncertain, especially if pop.~B is much more extended (as suggested by \citealt{Sacco2014}). The uncertainty on the centroid of pop.~B might further relax the request of partial line-of-sight superposition. 

\section{Conclusions}~\label{sec:conclude}
Several hints indicate that YSCs form from the merger of sub-clusters, born from the same molecular cloud. In this paper, we have simulated the collision between two sub-clusters. We have shown that the collision product reproduces some interesting features of the Gamma Velorum cluster, recently observed by the GES. The GES data (\citealt{Jeffries2014}) indicate that the Gamma Velorum cluster is composed of two kinematically decoupled populations: pop.~A and pop.~B. The two populations have a marginally different radial concentration and slightly offset centroids (by $\sim{}0.5$ pc).

Our simulations can naturally explain the RV offset between the two populations of the Gamma Velorum cluster, as well as their different intrinsic velocity dispersions (see Figure~\ref{fig:fig2}). Our simulations  also account for the different concentration and the different centroid of the two populations (Figures ~\ref{fig:fig4} and ~\ref{fig:fig5}). The GES data suggest that pop.~A is older by $1-2$ Myr, based on the Li depletion. Our simulations are consistent with a small age difference between the two populations.  In the scenario of a collision between two sub-clusters, the two populations can either be perfectly coeval or have formed in two slightly different star formation episodes. The latter hypothesis is supported by observations of similar age differences in other star forming regions (even if the question of age spread in star forming regions is still debated, see \citealt{Jeffries2011} and references therein). 

 We predict that the dispersion of the distribution of proper motions is broader for pop.~B than for pop.~A, even if the relative velocity vector between the two sub-clusters is aligned with the line-of-sight, as a result of the fact that pop.~B is supervirial (Figure ~\ref{fig:fig3}). Furthermore, if a component of the relative velocity vector is normal to the line-of-sight, we expect a difference in the average proper motions of pop.~A and pop.~B (e.g. Figure ~\ref{fig:fig7}). This will soon be tested with Gaia data. 


Furthermore, our simulations can help us reducing the allowed parameter space for the initial conditions of the Gamma Velorum cluster. For example, models with a low binary fraction ($f_{\rm bin}\sim{}0$) can hardly account for the observed RV data. Similarly, models in which the initial relative velocity between the clusters is $\Delta{}$RV$>3$ km s$^{-1}$ are in disagreement with the observed RVs. Star cluster models with an initial total mass $\le{}150$ M$_\odot$ can reproduce most of the observed  features, but cannot account for the formation of the massive $\gamma{}^2$ Velorum binary system. In contrast, models with a larger mass ($\ge{}500$ M$_\odot$) can host massive binary systems, similar to the $\gamma{}^2$ Velorum binary. 

Finally, the simulations suggest that pop.~B is not in virial equilibrium: supervirial models (with virial ratio $2<Q_{\rm B}<12.5$) are in better agreement with the GES data.  The virial ratio $Q$ is an essential ingredient of our models, and strongly affects the results: it is impossible to match the observed RV distribution of the Gamma Velorum cluster, without allowing $Q_{\rm B}$ to assume a value $>>0.5$. This result might indicate that the gas from the parent molecular cloud evaporated very fast, leaving the stellar population out of virial equilibrium. Thus, pop.~B is about to dissolve in the field. Furthermore, the physical meaning of $Q$ is connected with the star formation efficiency in the cloud. A larger value of $Q$ corresponds to a lower star formation efficiency, and consequently to a stronger effect of gas expulsion (e.g. \citealt{Tutukov1978,Hills1980,Lada1984,Baumgardt2007}). Thus, we can infer that pop.~A formed with higher star-formation efficiency, managed to survive gas evaporation, and preserved virial equilibrium, while pop.~B was formed in a less efficient star formation episode, and is not going to survive gas evaporation. A possible scenario is that pop.~B formed from a less dense molecular cloud core, where  star formation efficiency was intrinsically lower.
 Another possible scenario is that pop.~B formed later than pop.~A (as indicated by the Li depletion, \citealt{Jeffries2014}), and that the gas surrounding pop.~B was evaporated by the radiation field of pop.~A, quenching the star formation episode of pop.~B quite abruptly. 

Since it is supervirial, pop.~B expands faster than pop.~A. Simulated stars belonging to pop.~B lie at $>10$ pc from the centre of pop.~A at $t\gtrsim{}4$ Myr since the beginning of the simulation. \citet{Sacco2014} have recently found that some stars located in the NGC~2547 cluster have the same properties as those of pop.~B. Since the NGC~2547 cluster lies about two degrees ($\sim{}10$ pc) south of $\gamma{}^2$ Velorum, this result supports our scenario and strengthens the evidence that pop.~B is strongly supervirial.








The main intrinsic limits of our simulations  are the following: we have not included the parent molecular gas (the effects of gas evaporation are indirectly modelled by assuming a virial ratio $>0.5$), and we do not account for the Galactic tidal field directly. We will refine our simulations, by including the missing ingredients, in a forthcoming paper. Our model requires a fine-tuning: we need to assume that most of the RV offset between the two sub-clusters is aligned along the line-of-sight ($\pm{}12^\circ$; for larger angles, the distance between the centroids differs by $>3$ $\sigma{}$ with respect to the measured value). If the direction of the relative velocity vector between the two sub-clusters is significantly different from the line-of-sight, then we would expect to see a larger spatial offset between the two populations. If the Gamma Velorum cluster represents the case of a fortuitous alignment between velocity offset and line-of-sight, then we would expect to observe a  number of `twin' sub-clusters in young star formation complexes, showing a larger spatial offset than  the Gamma Velorum cluster. The forthcoming data by the Gaia mission will likely shed light on this issue, by providing accurate proper motion and parallax measurements.



Our results show that the Gamma Velorum cluster is an ideal test-bed to check different scenarios for the formation of young star clusters in the Milky Way.  The RV precision achieved by the GES allowed us to make an unprecedented comparison between the kinematics of simulated and observed clusters. Similar kinematic data of other young star clusters and associations in the GES sample will be essential to shed light on the formation of star clusters in the local Universe. 
 
\section*{Acknowledgments}
We thank the anonymous referee for the careful reading of the manuscript and for the helpful comments.
Based on data products from observations made with ESO Telescopes at the La Silla Paranal Observatory under programme ID 188.B-3002. These data products have been processed by the Cambridge Astronomy Survey Unit (CASU) at the Institute of Astronomy, University of Cambridge, and by the FLAMES/UVES reduction team at INAF/Osservatorio Astrofisico di Arcetri. These data have been obtained from the Gaia-ESO Survey Data Archive, prepared and hosted by the Wide Field Astronomy Unit, Institute for Astronomy, University of Edinburgh, which is funded by the UK Science and Technology Facilities Council. 

This work was partly supported by the European Union FP7 programme through ERC grant number 320360 and by the Leverhulme Trust through grant RPG-2012-541. We acknowledge the support from INAF and Ministero dell' Istruzione, dell' Universit\`a' e della Ricerca (MIUR) in the form of the grant "Premiale VLT 2012". The results presented here benefit from discussions held during the Gaia-ESO workshops and conferences supported by the ESF (European Science Foundation) through the GREAT Research Network Programme.

MM acknowledges financial support from the Italian Ministry of Education, University and Research (MIUR) through grant FIRB 2012 RBFR12PM1F, from INAF through grant PRIN-2011-1 and from CONACyT through grant 169554. The authors acknowledge the CINECA Award N. HP10B3BJEW, HP10CLI3BX, HP10CXB7O8, HP10C894X7, HP10CGUBV0, HP10CP6XSO and HP10C3ANJY for the availability of high performance computing resources and support.

\end{document}